\documentstyle[psfig]{mn}
\title{Basic parameters of three star clusters in the Small Magellanic Cloud: Kron~11, Kron~63 and NGC~121}
\author[Baume  at al.]
{G. Baume$^1$,\thanks{Member of Carrera del Investigador CONICET, Argentina.}
 N.E.D. No\"el$^2$,
 E. Costa$^3$,
 G. Carraro$^4$,\thanks{On leave from Dipartimento di Astronomia, Universit\`a di Padova,
 Vicolo Osservatorio 2, I-35122, Padova, Italy} and
 R.A. M\'endez$^3$ and M.H. Pedreros$^5$
 \thanks{email: gbaume@fcaglp.fcaglp.unlp.edu.ar (GB),
                noelia@iac.es (NN)
                costa@das.uchile.cl (EC),
                gcarraro@eso.org (GC),
                rmendez@das.uchile.cl (RAM),
                mpedrero@uta.cl(MP)} \\
      $^1$Facultad de Ciencias Astron\'omicas y Geof\'{\i}sicas (UNLP),
          Instituto de Astrof\'{\i}sica de La Plata (CONICET, UNLP), \\
          \hspace{0.2cm}Paseo del Bosque s/n, La Plata, Argentina \\
      $^2$Instituto de Astrof\'{\i}sica de Canarias, Tenerife 38200, Canary Islands, Spain \\
      $^3$Departamento de Astronom\'{\i}a, Universidad de Chile, Casilla 36-D, Santiago, Chile \\
      $^4$ESO, Alonso de Cordova 3107, Vitacura, Santiago de Chile, Chile \\
      $^5$Departamento de F\'{\i}sica, Facultad de Ciencias, Universidad de Taracap\'a, Casilla 7-D, Arica, Chile \\
}

\date{\it Submitted: *** 2008}
\pubyear{2005}
\begin{document}
\maketitle
\title{Star clusters in the SMC}

\begin{abstract}

We present observations for three star clusters, Kron~11, Kron~63 and NGC~121, in
the Small Magellanic Cloud. We have studied their structure and derived their
fundamental parameters by means of their luminosity functions, their color magnitude
diagrams and the Padova suite of isochrones. NGC~121 is a well studied object, for
which we confirm previous evidence about its old age and low metal content, and have
found that it is undergoing mass segregation. Kron~11 and Kron~63 are poorly populated
clusters which had never been studied so far. Kron~11 is several gigayears younger than
NGC~121, while Kron~63 is basically a very young star aggregate.  Both clusters are
immersed in dense stellar fields which share the same population properties, suggesting
that in their cases, cluster ages are consistent with typical ages of field stars.

\end{abstract}


\section{Introduction}

Star clusters are widely considered as single-age and single-metallicity stellar
populations. This has allowed to use them as stellar population tracers
to infer the Star Formation History (SFH) of their parent galaxy. In particular,
star clusters in the Magellanic Clouds (MC) have become a challenging domain for
stellar and galactic evolutionary models because these clusters differ significantly
from those found in our galaxy. These differences are commonly attributed to the
profoundly different MC environment and, hence, to a different chemical and dynamical
evolution.

In this paper we investigate three star clusters of the Small Magellanic Cloud (SMC):
Kron~11 (= Lindsay 20), Kron~63 (=Lindsay 88) and NGC~121 (=Lindsay 10).  We have
determined their fundamental parameters: morphology, age and metallicity. The observations
were secured as a part of a comprehensive study of the MC, which includes the study of
stellar clusters (see e.g. NGC~2154, Baume et al. 2007), the study of their global SFH
(No\"el et al. 2007, hereafter NGCM;  No\"el et al 2008) and the determination
of their absolute proper motions with respect to background QSOs (Pedreros et al. 2006,
Costa et al. 2008).

Kron~11 and Kron~63 had so far only been cataloged (see e.g. Kron 1956; Lindsay1958;
Bica \& Schmitt 1995), and only a rough estimation of their properties was available
(see Kontizas et al. 1990 and references therein). Here we provide the first derivation
of their basic properties from deep photometric observations.  Adding up information on
unstudied clusters certainly may help to better understand the age distribution of clusters
in the SMC, a subject which is quite disputed (see e.g. Mighell, Sarajedini \& French 1998a,b;
Rich et al. 2000; Chiosi et al. 2006; Gieles et al. 2007; Piatti et al. 2007).

NGC~121 is a very well studied cluster (see e.g. Glatt et al. 2008, and references therein).
It is the oldest SMC cluster and it is currently considered to be the only true
globular cluster in the SMC.  We expected that our observations of NGC~121 would
allow us to focus on conflictive aspects of previous observations of this cluster,
but, unfortunately, poor photometric conditions (see Sect.~4) prevented us form
reaching our original goal.

The layout of the paper is as follows.  In Sect.~2 we describe the observations, and
in Sect.~ 3 and 4 we describe the reduction procedure and the photometric completeness,
respectively. In Sect.~5 we discuss the cluster's morphologies. In Sect.~6 and 7 we
present their Luminosity Functions (LFs) and Color Magnitude Diagrams (CMDs), respectively.
In Sect.~8 we analyze each cluster individually and, finally, in Sect.~9 we summarize and
further discuss our results.

\section{Observations and Reductions}

\begin{figure*}
\centerline{\psfig{file=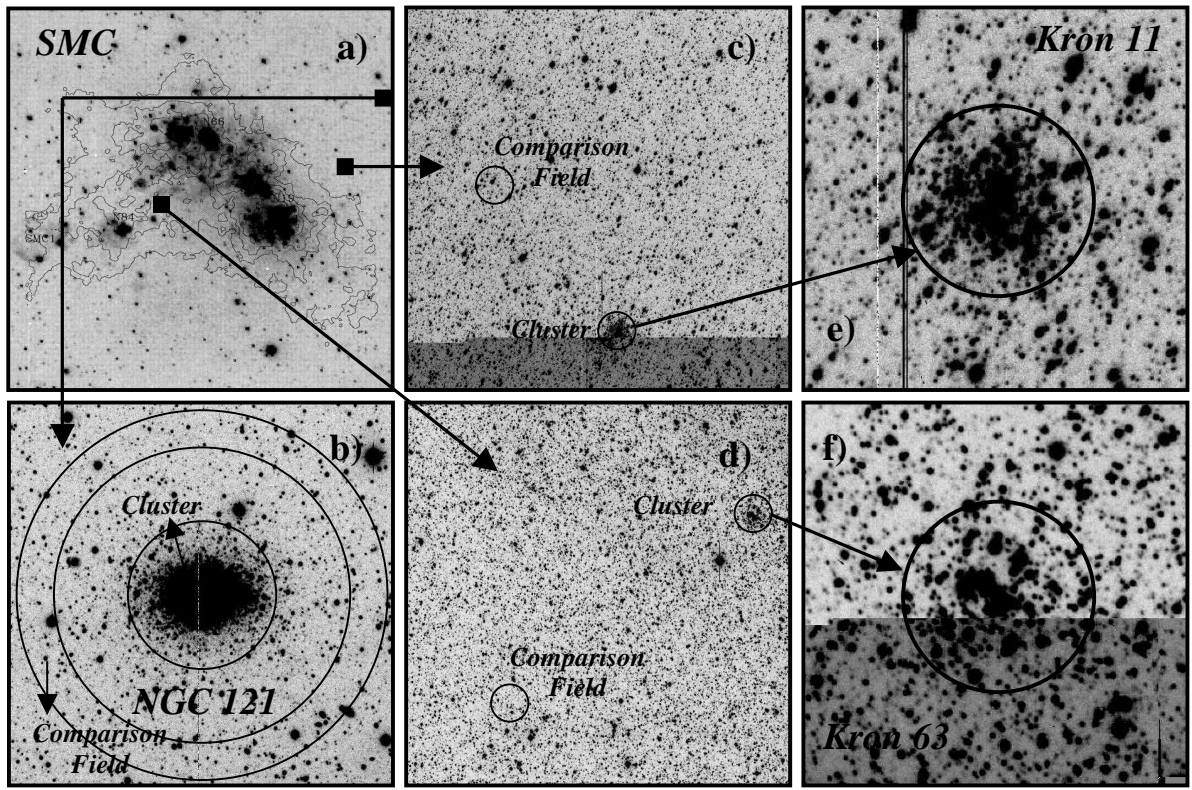,width=16cm}}
\caption{{\bf a)} $H\alpha$ image of the SMC from Kennicutt et al. (1995) with HI
contours (Stanimirovi\'c et al. 1999).  Black squares indicate the approximate
position of the regions observed.
{\bf b), c)} and {\bf d)} $R$ band images of the specific fields studied in the
regions of NGC~121, Kron~11 and Kron~63, respectively.  The size of the regions is
$8\farcm85 \times 8\farcm85$.  Circles indicate the adopted {\it Cluster} areas and
{\it Comparison Fields} in each case (see Sect.~6 and 7) {\bf e)} and {\bf f)} Expanded
views of Kron~11 and Kron~63, respectively. In all cases North is up and East to the left.}
\label{fig:xy}
\end{figure*}

$B(R)_{KC}$ images of the NGC~121, Kron~11 and Kron~63 regions of the SMC were acquired
with a 24$\mu$ pixels Tektronix 2048$\times$2048 detector attached to the Cassegrain
focus of the du~Pont 2.5-meter telescope (C100) at Las Campanas Observatory, Chile.
Gain and read noise were 3 e-/ADU and 7 e-, respectively.  This set-up provides
direct imaging over a field of $8\farcm85 \times 8\farcm85$ with a scale of
$0\farcs259$/pix. The relatively large field of view allowed us to include the cluster
and a good sample of the adjacent SMC field in each case. The $B$ and $R$ bandpasses were selected
to satisfy the needs of both the astrometry and photometry programs. The fields of interest
are shown in Fig.~\ref{fig:xy}.

Details on the observations available for NGC~121 are given in Table~1 of this paper,
whereas details on the observations available for Kron~11 and Kron~63 are given in
Table~1 of NGCM: fields qj0036 (Kron~11) and qj0111 (Kron~63). Typical FWHM of the
Kron~11 and Kron~63 data is about $0\farcs9$, whereas a much larger value
($\approx 1\farcs8$) resulted for the NGC~121 data.

All frames were pre-processed in a standard way using the IRAF\footnote{IRAF is
distributed by NOAO, which is operated by AURA under cooperative agreement with the NSF.}
package CCDRED.  With this purpose, zero exposures and sky flats were taken every night.

\begin{table}
\begin{center}
\tabcolsep 0.5truecm
\caption{NGC~121 observations}
\begin{tabular}{lccr@{$\times$}l}
\hline
\multicolumn{1}{c}{Date} & Airmass & Filter & \multicolumn{2}{c}{Exp. Time} \\
                         &         &        & [sec. & N]                    \\
\hline
06-10-07 &  1.41 & $B$ & ~60 & ~1 \\
         &  1.45 & $B$ & 600 & ~9 \\
\hline
07-10-07 &  1.49 & $R$ & ~60 & ~7 \\
         &  1.52 & $R$ & 400 & 15 \\
         &  1.43 & $R$ & 600 & ~1 \\
         &  1.41 & $B$ & ~60 & ~7 \\
         &  1.45 & $B$ & 800 & ~1 \\
\hline
\end{tabular}
\begin{minipage}{6cm}
\vspace{0.1cm}
{\bf Notes:} $N$ indicates the number of frames obtained.
\end{minipage}
\end{center}
\end{table}

\section{The Photometry}

While the full reduction of the photometric data available for NGC~121 is
presented here, we refer to NGCM for any details about the reduction of the
Kron~11 and Kron~63 data. In fact, we limited ourselves to use that
data-set to derive the first estimates of the fundamental parameters
of these two clusters, which happened to fall inside stellar fields
observed for other purposes.

\subsection {Standard star photometry}

Our instrumental photometry was defined by the use of the Harris $UBVRI$ filter set,
which constitutes the default option at the C100 for broad-band photometry on the
standard Johnson-Kron-Cousins system.  On photometric nights, standard star areas
from the catalog of Landolt (1992) were observed multiple times to determine the
transformation equations relating our instrumental ($b$,$r$) magnitudes to the
standard ($B$,$R_{KC}$) system.  These standard star areas were selected to provide
a wide range in colors. A few of them were followed each night up to about 2.0
airmasses to determine atmospheric extinction optimally.  Aperture photometry was
carried out for all the standard stars using the IRAF PHOTCAL package.  To tie our
observations to the standard system, we used transformation equations of the form:\\

\noindent
$ b = B + b_1 + b_2 * X + b_3~(B-R)$ \\
$ r = R + r_1 + r_2 * X + r_3~(B-R)$ \\

\noindent
In these equations $b,r$ are the aperture magnitude already normalized to 1 sec,
and $X$ is the airmass. We did not include second-order color terms because they
turned out to be negligible in comparison to their uncertainties. The values of the
transformation coefficients for our NGC~121 observations are listed in Table~2,
whereas those for Kron~11 and Kron~63 can be found in Table 3 of NGCM.

\begin{table} 
\begin{center}
\tabcolsep 0.7truecm
\caption {Transformation Equations Coefficients (NGC~121 observations)}
\begin{tabular}{ll}
\hline
$b_1 = +1.058 \pm 0.025$ & $r_1 = +0.613 \pm 0.049$ \\
$b_2 = +0.213 \pm 0.019$ & $r_2 = +0.136 \pm 0.038$ \\
$b_3 = -0.043 \pm 0.004$ & $r_3 = -0.009 \pm 0.006$ \\
\hline
\end{tabular}
\end{center}
\end{table}

\subsection {Clusters and SMC field photometry}

We followed the procedure outlined in Baume et al. (2004). We first averaged
images taken on the same night, and with the same exposure time and filter, to
remove cosmic rays and improve the signal-to-noise ratio of the faintest stars.
Then, instrumental magnitudes and $X,Y$ coordinates were obtained using the Point Spread
Function (PSF) method (Stetson 1987). Finally, all data from different filters
and exposures were combined and calibrated using DAOMASTER (Stetson 1992).  The
photometry will be made available electronically at CDS.

\begin{figure}
\begin{center}
\centerline{\psfig{file=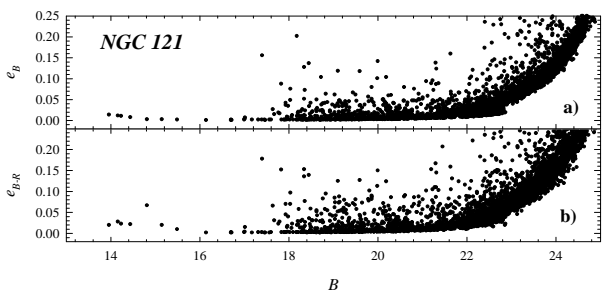,width=8cm}}
\caption{Photometric errors in $B$ and $B-R$, given by DAOPHOT and DAOMASTER,
plotted as a function of $B$ magnitude (NGC~121 observations).}
\label{fig:err}
\end{center}
\end{figure}

In Fig.~\ref{fig:err} we present the photometric errors trends (from DAOPHOT
and DAOMASTER) as a function of the $B$ magnitude for our NGC~121 observations.

\section{Photometric Completeness} \label{sec:comp}

In order to derive the LFs of each cluster, and to compare CMDs decontaminated of
field stars with isochrones, a completeness analysis of the photometric data is
required.


\subsection{Kron~11 and Kron~63}

The completeness analysis of the photometric data secured for fields {\it qj0036} and
{\it qj0111} of the SMC (see NGCM) -in which Kron~11 and Kron~63 are included-
was carried out by No\"el et al. (2008b), using the same technique we have applied
to NGC~121. Both the photometric extraction and the completeness results are
discussed there.

\subsection{NGC~121}

A completeness correction for the photometric data of NGC~121 was determined by
means of artificial-star experiments (see e.g. Baume et al. 2007). We added
a total of 40000 artificial stars in random positions (with the same luminosity
function as that of our real sample) to our images. To avoid the creation of
overcrowding, in each experiment we added the equivalent to only 15$\%$ of the
original number of stars. The completeness factor is defined as the ratio between
the number of artificial stars recovered and the number of artificial stars added.
These ratios were computed for different $B$ and $R$ magnitude bins (0.5 magnitude
wide).  They are listed in Table~3, both for the cluster (r $< 103\farcs6$) and
for a representative comparison field (see Sect.~6.1). To obtain the completeness
factor for a particular star with a given $B-R$ color, the corresponding factors
in $B$ and $R$ have to be multiplied.

\begin{table}
\tabcolsep 0.5truecm
\caption {Completeness analysis results for NGC~121}
\begin{tabular}{cc@{}cc@{}c}
\hline
$\Delta mag$ & \multicolumn{2}{c}{NGC~121}            & \multicolumn{2}{c}{Field}  \\
             & \multicolumn{2}{c}{(r $< 103\farcs6$)} &                            \\
             & $B$ mag. & ~$R$ mag.                    & $B$ mag. & ~$R$ mag.      \\
\hline
18.0-18.5    & 100.0\%  & ~100.0\%                    & 100.0\%  & ~100.0\%        \\
18.5-19.0    & 100.0\%  & ~~99.2\%                    & 100.0\%  & ~100.0\%        \\
19.0-19.5    & ~97.4\%  & ~~87.3\%                    & 100.0\%  & ~100.0\%        \\
19.5-20.0    & ~92.6\%  & ~~78.1\%                    & 100.0\%  & ~100.0\%        \\
20.0-20.5    & ~83.1\%  & ~~64.2\%                    & 100.0\%  & ~100.0\%        \\
20.5-21.0    & ~74.2\%  & ~~61.3\%                    & 100.0\%  & ~100.0\%        \\
21.0-21.5    & ~59.2\%  & ~~57.6\%                    & 100.0\%  & ~100.0\%        \\
21.5-22.0    & ~55.3\%  & ~~46.1\%                    & 100.0\%  & ~~95.2\%        \\
22.0-22.5    & ~56.4\%  & ~~32.7\%                    & 100.0\%  & ~~84.4\%        \\
22.5-23.0    & ~42.1\%  & ~~30.5\%                    & ~96.3\%  & ~~73.2\%        \\
23.0-23.5    & ~30.3\%  & ~~23.2\%                    & ~86.2\%  & ~~66.7\%        \\
23.5-24.0    & ~31.2\%  & ~~21.3\%                    & ~77.4\%  & ~~56.5\%        \\
24.0-24.5    & ~22.7\%  & ~~15.2\%                    & ~62.1\%  & ~~42.7\%        \\
24.5-25.0    & ~23.9\%  & ~~13.7\%                    & ~53.1\%  & ~~31.9\%        \\
25.0-25.5    & ~16.5\%  & ~~10.4\%                    & ~40.4\%  & ~~19.6\%        \\
25.5-26.0    & ~12.1\%  & ~~~7.2\%                    & ~33.2\%  & ~~14.4\%        \\
26.0-26.5    & ~~8.2\%  &                             & ~20.3\%  &                 \\
\hline
\end{tabular}
\end{table}

\section{Radial Density and Surface Brightness Profiles}

With the aim of deriving the clusters extension and shape, as a first step we
estimated the position of their centers searching for the highest peak in the stellar
density. This was done by visual inspection of the available images. The coordinates
of the cluster's centers are given in Table~4; they turned out to be very similar to
those given in the $SIMBAD$ database.

\begin{table*} 
\tabcolsep 0.5truecm
\caption {Main properties of the studied star clusters}
\begin{tabular}{lccccccc}
\hline
Object  & X    & $\alpha_{2000}$ & Radius                 & $V_O - M_V$ & z             \\
        & Y    & $\delta_{2000}$ &                        & $E(B-V)$    & Log(age[yr]) \\
\hline
Kron~11 & 1166 & ~00:36:23.6     & $25\farcs-30\farcs$    & 18.7        & 0.0010        \\
        & 1285 & -72:28:39.0     &                        & 0.05        & 9.70 - 9.80   \\
\hline
Kron~63 & 1850 & ~01:10:47.3     & $20\farcs-25\farcs$    & 18.7        & 0.0006        \\
        & ~615 & -72:47:34.4     &                        & 0.07        & 7.75 - 8.00   \\
\hline
NGC~121 & 1030 & ~00:26:48.2     & $70\farcs - 200\farcs$ & 19.1        & 0.0006        \\
        & 1025 & -71:32:07.3     &                        & 0.04        & 10.0 - 10.1   \\
\hline
\end{tabular}
\begin{minipage}{12.5cm}
\end{minipage}
\end{table*}

We then compute the cluster's size, which was done constructing radial profiles using
two methods (see Baume et al 2007): the radial stellar density profile and the radial
surface brightness profile. In the first method, stars are counted in a number of successive
concentric rings around the adopted cluster center, and then a division is made by the
respective areas.  In the second method, the flux ($ADUs/area$) within concentric annuli
is measured directly on the cluster images.

A measure of the cluster's radius was obtained by fitting an EEF model (see Elson et al.
1987), appropriate for MC clusters (Mackey \& Gilmore 2003). The expression used was:\\

\begin{figure}
\begin{center}
\centerline{\psfig{file=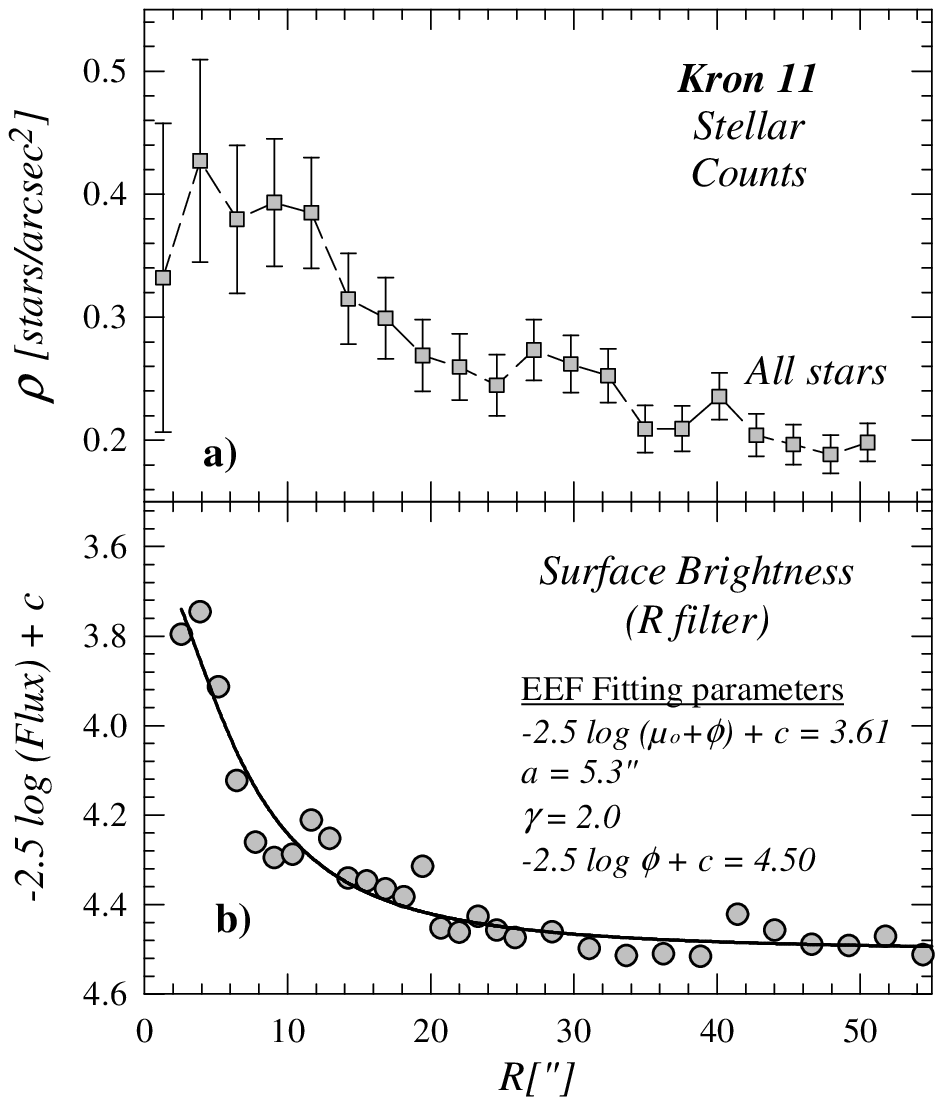,width=7.5cm}}
\centerline{\psfig{file=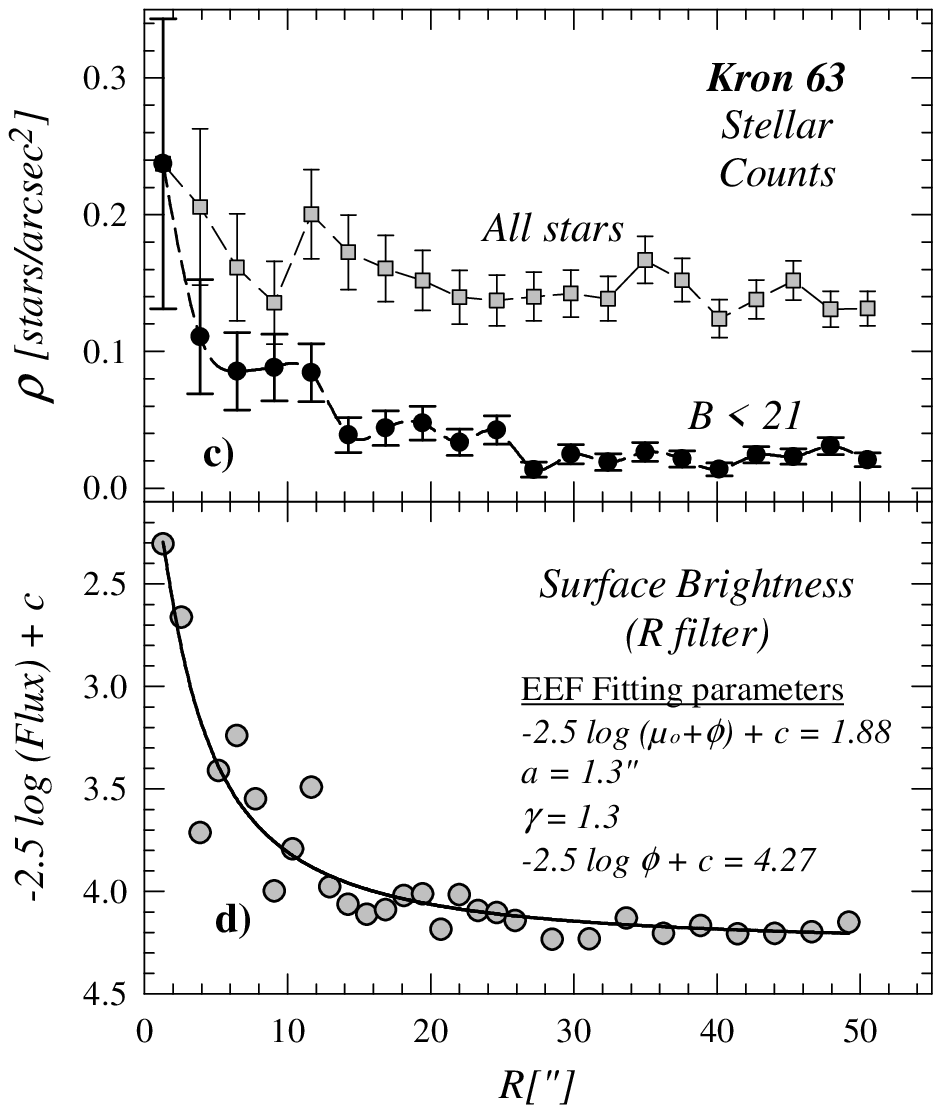,width=7.5cm}}
\caption{Radial profiles for Kron~11 and Kron~63.
{\bf a)} and {\bf c)} Radial stellar density profiles.
{\bf b)} and {\bf d)} Radial flux profiles (grey circles) and the EFF model fits
(solid curves) together with the computed fitted parameters.  $c$ is an arbitrary
constant with a different value for each cluster)}
\label{fig:rad-kron}
\end{center}
\end{figure}

\begin{figure}
\begin{center}
\centerline{\psfig{file=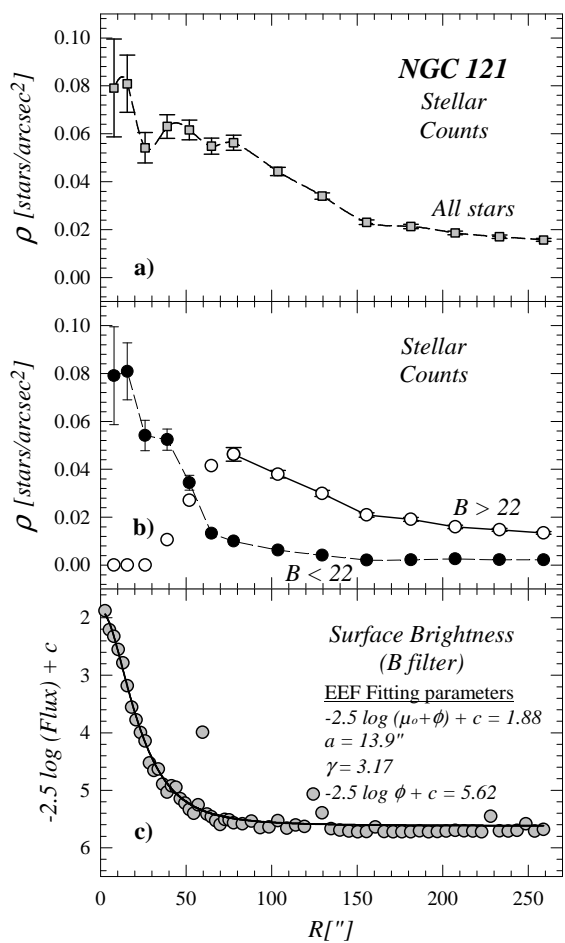,width=7.5cm}}
\caption{Radial profiles for NGC~121. {\bf a)}Radial stellar density profiles.
{\bf b)} Radial stellar density profiles for two different magnitude intervals.
{\bf c)}Radial flux profile (grey circles) and the EFF model fit (solid curve)
together with the computed fitted parameters. $c$ is an arbitrary constant)}
\label{fig:rad-ngc}
\end{center}
\end{figure}

\noindent
$\mu = \mu_0 [1 + (r/a)^2]^{-\gamma/2} + \phi$ \\

\noindent
where $r$ is the distance from the adopted cluster center, $\mu_o$ the central
surface brightness, $a$ is a measure of the core radius, $\gamma$ the power
law slope at large radii, and, finally, $\phi$ is the field surface brightness.
The resulting profiles together with the computed parameters are presented in
Figs.~\ref{fig:rad-kron} and \ref{fig:rad-ngc}.

\section{Observed Luminosity Functions}

\begin{figure*}
\begin{center}
\begin{tabular}{ccc}
\psfig{file=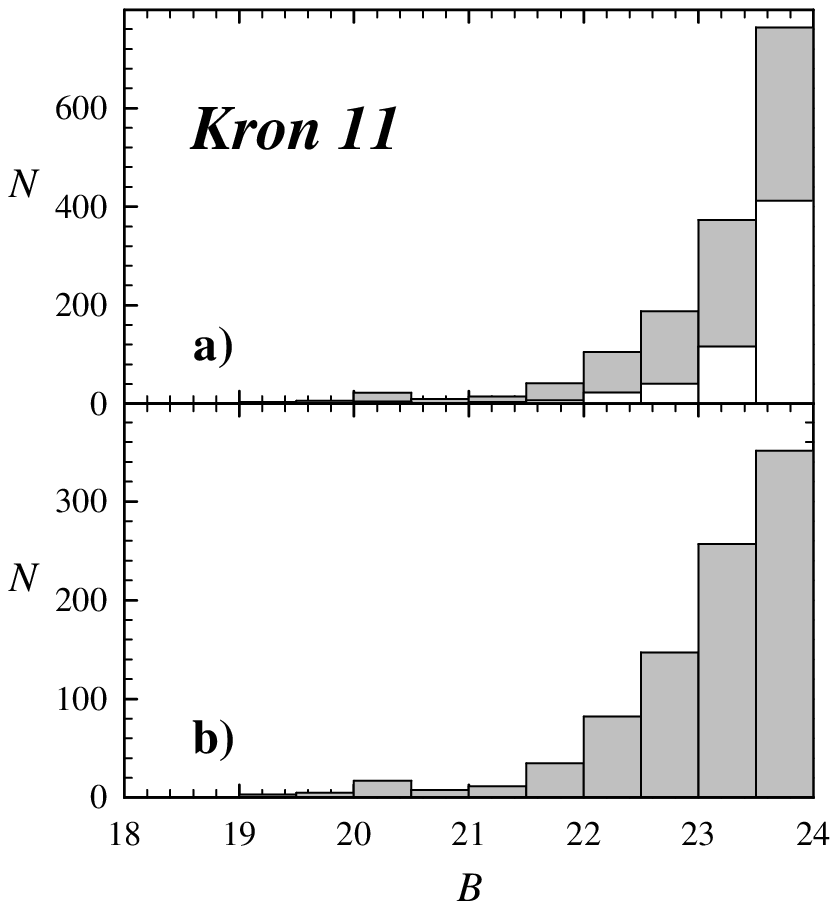,width=5.6cm} &
\psfig{file=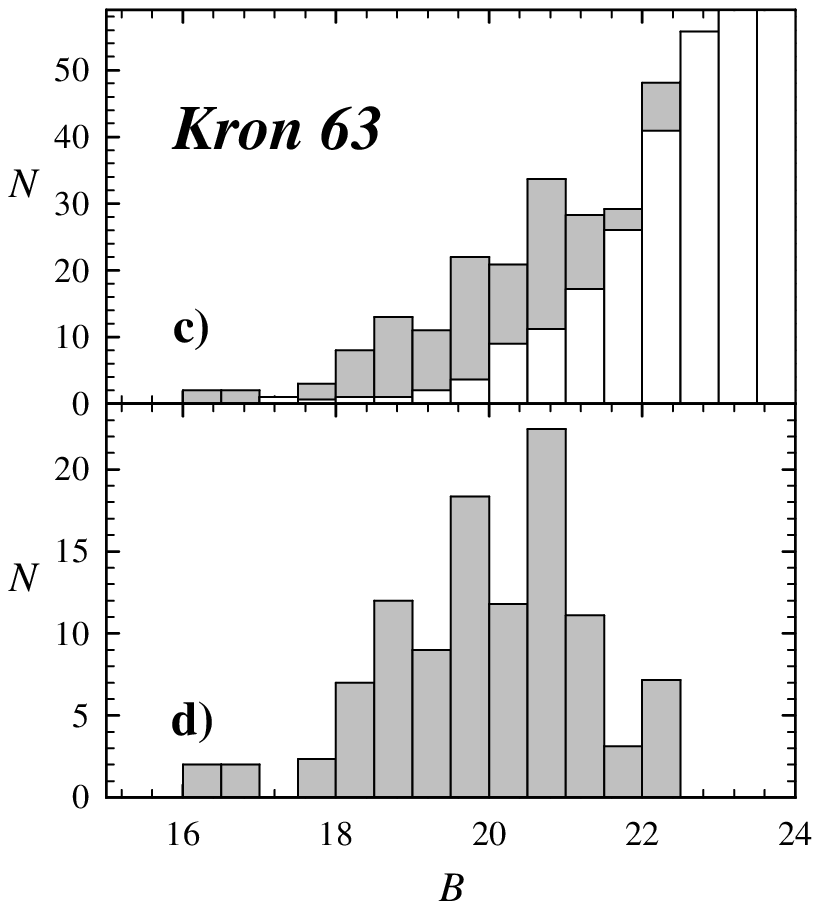,width=5.6cm} &
\psfig{file=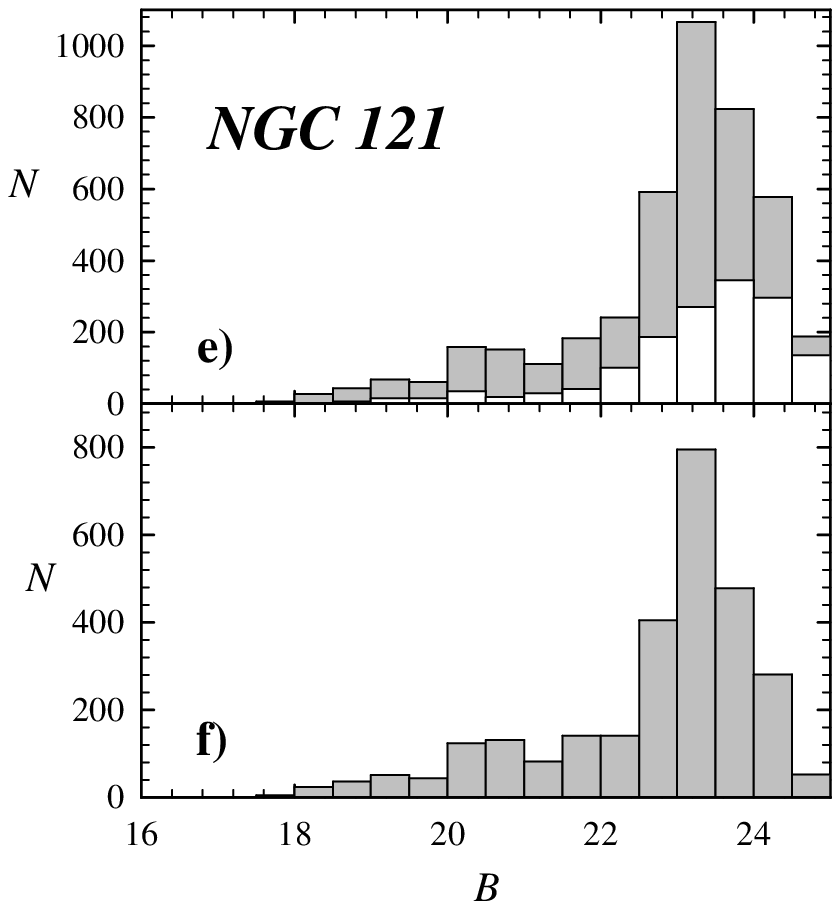,width=5.6cm} \\
\end{tabular}
\caption{{\bf a)}, {\bf c)} and {\bf e)} Luminosity functions corrected by completeness factors (see
Sect.~\ref{sec:comp}). Grey histogram corresponds to the {\it Cluster} areas, whereas the white
histogram corresponds to the respective {\it Comparison Field} areas.
{\bf b)}, {\bf d)} and {\bf f)} Luminosity functions obtained for each cluster after the correction
by field contamination}
\label{fig:lfs}
\end{center}
\end{figure*}

{\bf To obtain the observed LFs of the clusters, we first counted the stars in each
apparent magnitude bin, both in the {\it Cluster} area and in their
respective {\it Comparison Field} (See Fig.~\ref{fig:xy}). In each case they were
corrected by the completeness factors given in Sect.~\ref{sec:comp}. The resulting
histograms are shown in the upper panels of Fig.~\ref{fig:lfs}.  A cleaner picture
of the cluster's population is obtained by substracting these histograms in the
sense {\it Cluster - Field}. The results are shown in the lower panels of Fig.~\ref{fig:lfs}.}

\section{Color-Magnitude Diagrams}

\begin{figure*}
\begin{center}
\centerline{\psfig{file=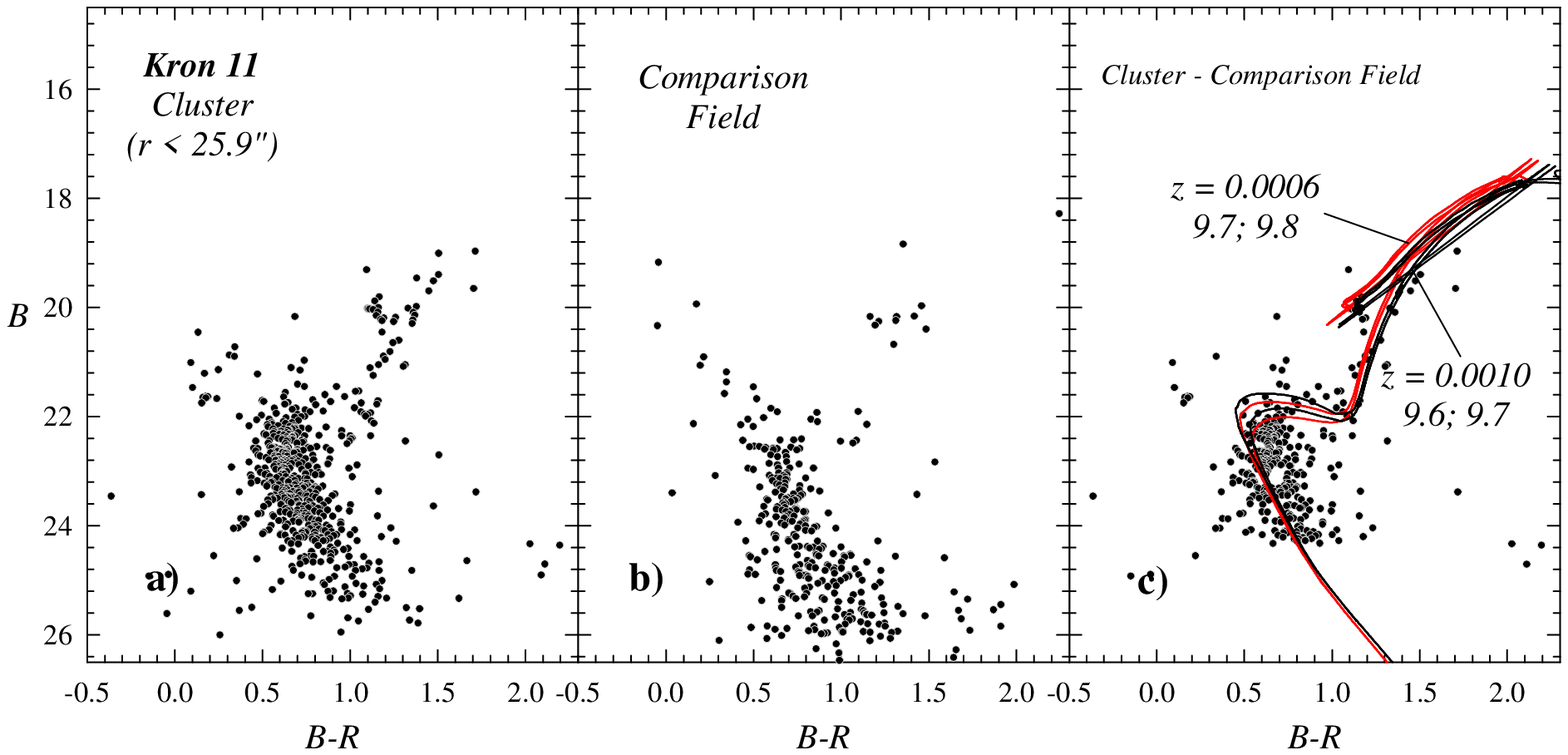,width=16cm}}
\centerline{\psfig{file=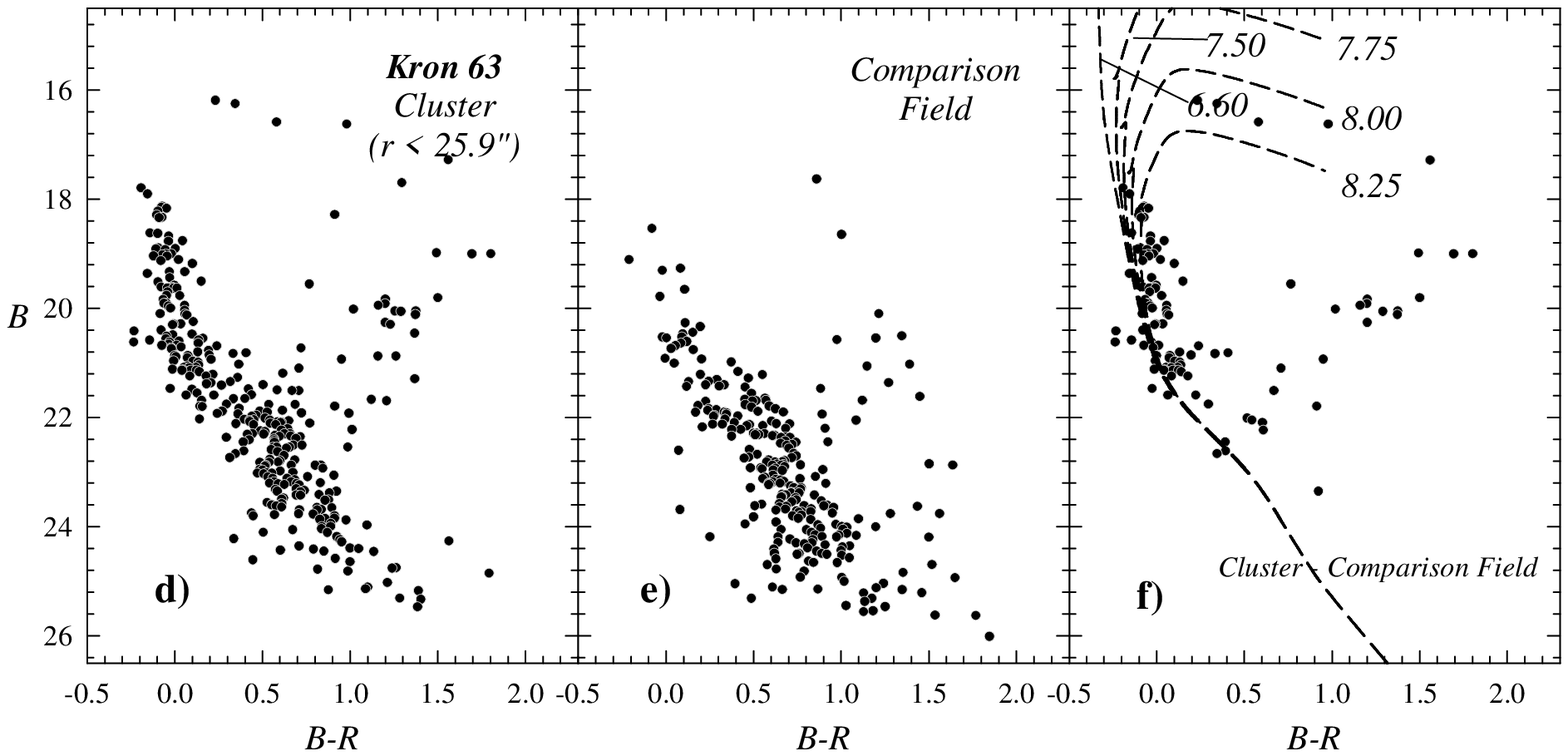,width=16cm}}
\caption{CMDs of Kron~11 and Kron~63. Left panels ({\bf a)} and {\bf d)}) are the CMDs of stars
placed in the $Cluster$ areas. Central panels ({\bf b) and {\bf e)}} are the CMDs of stars in the
respective comparison field. Right panels ({\bf c) and {\bf f)}} present the statistical difference
between the other panels ($Cluster~area$ - $Comparison~Field$). Solid curves are the best
fitting isochrones from Girardi et al. (2000). Numbers indicate Log(age[yr]) and adopted metallicities}
\label{fig:cmds-kron}
\end{center}
\end{figure*}

\begin{figure*}
\begin{center}
\centerline{\psfig{file=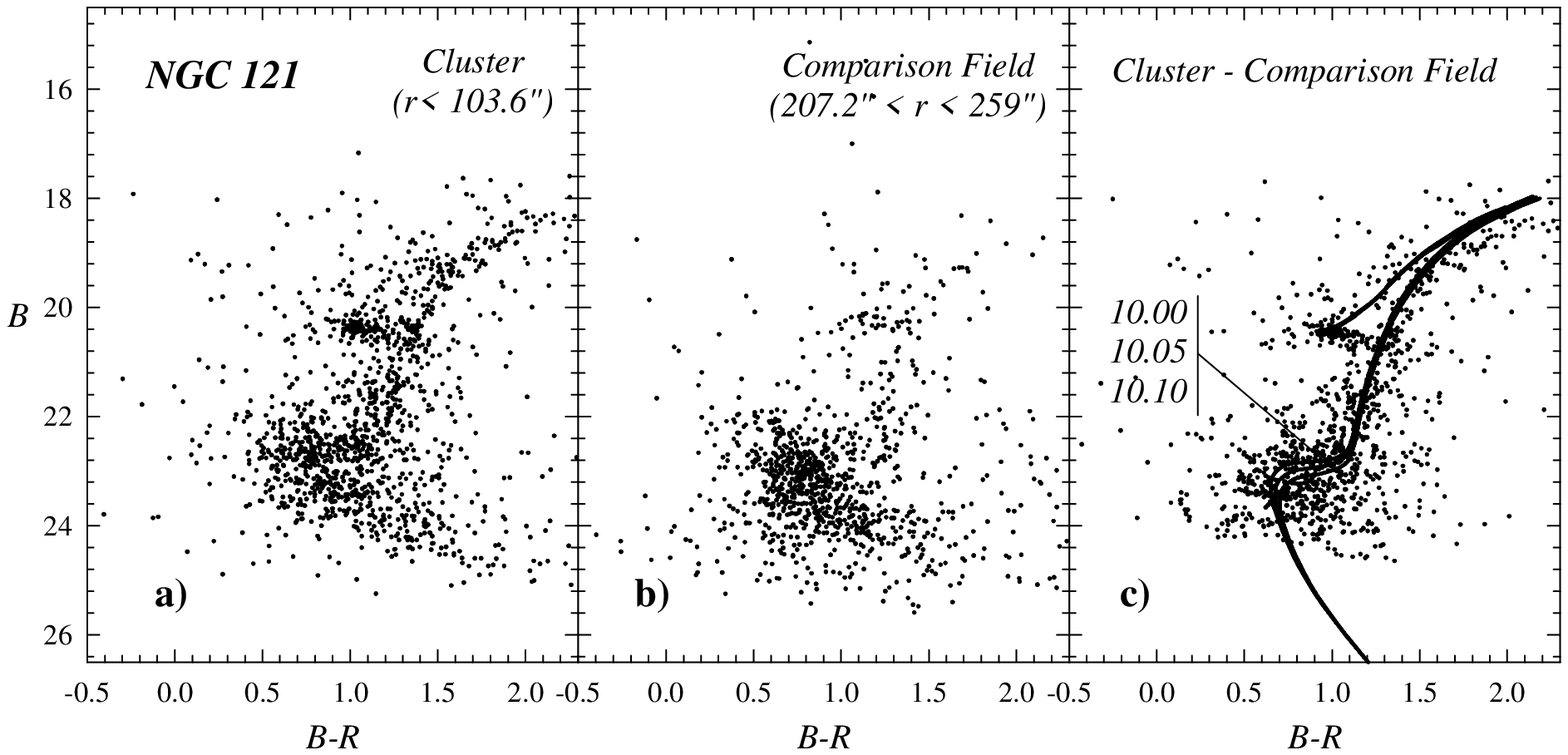,width=16cm}}
\caption{{\bf} CMDs of NGC~121 observations. Panels have the same meaning as in
Fig.~\ref{fig:cmds-kron}}
\label{fig:cmds-ngc}
\end{center}
\end{figure*}

{\bf In Fig.~\ref{fig:cmds-kron} we present CMDs for Kron~11 and Kron~63, and in
Fig.~\ref{fig:cmds-ngc} we present CMDs for NGC~121. Left panels show CMDs for
stars measured in the selected areas indicated as {\it Cluster}, while the central panels
show CMDs for stars measured in the selected areas indicated as {\it Comparison Field}
in Fig.~\ref{fig:xy}. These latter were chosen at appropriate distances from the cluster
centers to minimize the presence of cluster stars in them. These figures show that the
{\it Cluster} CMDs suffer an important contamination by SMC field stars.

To obtain better estimates of clusters characteristics we applied the statistical
decontamination method of the CMDs described in Vallenari et al. (1992) and
Gallart et al. (2003), and used in our study of NGC~2154 (Baume et al. 2007). In
this procedure, a statistical subtraction of field stars is carried out making a
star-by-star comparison between the selected {\it Comparison Field} and the {\it Cluster}
area. Briefly, for any given star in the {\it Comparison Field} we search for the most
similar (in color and magnitude) star in the {\it Cluster} area, and remove it from the
corresponding CMD. It should be noted that this procedure takes into account the difference
in completeness level between the {\it Cluster} area and their {\it Comparison Field}, and
also the increase of the photometric error with increasing magnitude (the search ellipse is
a function of the magnitude to take into account this latter effect). Figs.~\ref{fig:cmds-kron}c,
\ref{fig:cmds-kron}f and ~\ref{fig:cmds-ngc}c present the resulting decontaminated CMDs.}

Quick inspection of Figs. ~\ref{fig:cmds-kron} and \ref{fig:cmds-ngc} shows that our
observations of NGC~121 are clearly less deep than those of Kron~11 and Kron~63.
Comparison of the data presented in Table 1 of this work, with that presented in
Table 1 of NGCM shows that although total exposure time is a factor, the main reason for
this difference is the worse seeing conditions seen throughout our NGC~121 observing
runs ($\approx 1\farcs8$), amplified by crowding in the much denser NGC~121 field.

Interestingly, the three clusters share the same stellar population features of the SMC
field they are immersed in.  NGC~121 and Kron~11 are located in old fields, and look old,
whereas Kron~63 is located in a younger field, and appears to be young. It should be noted
that these stellar populations are typical of the North-West and South-East regions of
the SMC (see NGCM for details).

To derive the cluster's basic parameters (age, metallicity) we compared their CMDs
with theoretical isochrones from the Padova stellar evolution models (Girardi et al.
2000). The selection of the metallicities was done choosing those isochrones that produce
the best fit to the red giant branch (RGB) -when present- and to the main sequence (MS).
To perform the above comparisons it was necessary to adopt a distance modulus and to
assume a mean value for the reddening of each cluster. These values are indicated in
Table~4 and they are consistent with several previous estimates (see next sections).
In all cases, the normal relations: $E(B-R) = 1.57~E(B-V)$ and
$B-M_B = (V_O - M_V) + 4.1~E(B-V)$, have been adopted to calculate the color excess
and the apparent distance modulus, respectively.

\section{Individual Cluster Analysis}

\subsection{Kron~11}

The radial profiles of this object (Figs.~\ref{fig:rad-kron}a and b) reveal the clear presence
of a cluster above the stellar field up to a distance of about 25-30$^{\prime\prime}$ from the
adopted center. The surface brightness profile presents however a moderate spread around the
fitted curve of the EEF radial model. This fact is due the poor radial symmetry of the object
(see Fig.~\ref{fig:xy}e).

As previously noted, the CMDs of this cluster and its {\it Comparison~Field}
(Figs.~\ref{fig:cmds-kron}a and b), do not show significant differences between them, in
the sense that they are apparently composed by similar stellar populations. However, according
to its corrected LF (see Fig.~\ref{fig:lfs}b), the cluster produces a clear over-density
of stars corresponding to its MS and a horizontal branch (HB). Test fits with several isochrones
of different age and metallicity indicate that this cluster has a slightly higher metallicity
than that normally adopted for the SMC (Maeder et al. 1999).

With an age of $\sim$ 6 Gyr, Kron~11 sums up to the intermediate age group of SMC star
clusters which fall in the LMC age gap (Girardi et al. 1995). Although the percentage of star
clusters in this age range is still poorly constrained (Rafelski \& Zarisky 2005), the present
result points out to a clear difference in the rate of cluster formation in between the MCs.

\subsection{Kron~63}

This cluster appears as a weak concentration of bright stars against the rich SMC field.
Before analyzing the properties of this overdensity, we can test its reality by means
of a {\it t-Student} test.  A comparison of its corrected LF (see Fig.~\ref{fig:lfs}d)
with typical stellar fluctuations present in the rest of the observed region shows
that the Kron~63 "fluctuaction" is four times larger than the average stellar field
fluctuations. Applying these values to a {\it t-Student} test, we obtain a nearly 100\%
of reliability, meaning that Kron~63 is a real star cluster, and not just a statistical
fluctuation of the field population.

Examination of the radial profile of this object (Fig.~\ref{fig:rad-kron}c), shows that if all
detected stars are taken into account it does not stand out clearly over the field, but, if only
the brightest stars are considered ($B < 21$), the cluster becomes more prominent and we find
that it is composed mainly of blue stars ($B-R < 0.5$). The surface brightness profile of
this object also presents an important spread around the fitted curve of the EEF radial model
(see Fig.~\ref{fig:rad-kron}d).  Both radial profiles indicate however that this cluster extends
up to a distance of about 20-25$^{\prime\prime}$ from the adopted center.

The CMDs of this cluster (Figs.~\ref{fig:cmds-kron}c and d) exhibit characteristics typical of
an object much younger than Kron~11; it presents a significant MS population in an ample range
of magnitudes, and it does not seem to posses luminous red stars. Therefore, the best isochrone
was found on the basis of the MS morphology, and adopting a metallicity considered normal for
the SMC (see Table~4).

We notice that after the decontamination process the MS of Kron ~63 shortens significantly
compared to that of Kron 11.  This can be explained by the difference in mass between the
two clusters.  As shown by the radial density profiles derived in Sect.~5. (and by Fig.~1),
although significatively older, Kron~11 still looks more compact and massive than Kron~63.
At birth Kron~11 must have been more massive than Kron~63, and, having a larger potential
well, it has survived longer and managed to retain a larger number of low mass stars.
Low mass star-depleted MSs are not rare, and are often found, at different ages, in
Milky Way star clusters (Patat \& Carraro 1995, Bica et al. 2001, Nordstrom et al. 1997,
de la Fuente Marcos et al. 2008).


\subsection{NGC~121}

The radial profile based on stellar counts (Fig.~\ref{fig:rad-ngc}a), clearly shows the
limitations of the data available for this cluster due to poor seeing conditions, as
mentioned in Sect. 5. This is reflected in the very low stellar density values obtained for
this object. It is still possible however, to describe some cluster properties by selecting
appropriate magnitude intervals (Fig.~\ref{fig:rad-ngc}b), and by means of the surface
brightness profile (Fig.~\ref{fig:rad-ngc}c).  Both plots reveal the following issues:\\

\begin{itemize}
\item The adopted EEF profile fits nicely to the data, indicating this cluster is very
      well represented by this model.
\item The total cluster brightness is dominated by the brightest stars and the cluster is
      apparently extending up to near 70$^{\prime\prime}$ of the adopted center.
\item If only the faintest stars are included ($B > 22$), then the radial extension can reach
      up to 200$^{\prime\prime}$ from that center. This difference in radius can be interpreted
      as a product of the relaxation process suffered by the cluster, causing that the massive
      (brightest) stars move to its center and low-mass (faintest) stars to its outskirts.
\end{itemize}

\noindent
Due to peculiarities in the distribution of field stars across the cluster region,
a very efficient decontamination of the CMD of NGC~121 was not possible. Inspection of
Fig.~\ref{fig:cmds-ngc}c shows that in the clean CMD there are several groups of stars
that are more numerous than in the field, despite the fact that both regions have the
same area. In the clean CMD some field stars still remain above the MS, to the left of
the MS and everywhere in the red clump region. This effect surely results from the
statistically low number of stars in those regions of the CMD. Nonetheless, the procedure
was very effective and helped us to improve the shape of the turnoff region (TO), and to
remove several field RGB and HB stars.

Because there are good (and deeper) previous studies of this object (see e.g. Glatt et
al. 2008, and references therein), here we limit ourselves to check that their derived
parameters are compatible with our photometry. We achieved the best fit using the parameters
given in Table~4.  As found by Glatt et al. (2008), we highlight the notable difference between
theoretical models and observations in the upper RGB region.  Model predictions are systematically
bluer and more luminous than the observed RGBs, irrespective of the adopted passband.

\section{Conclusions}

We present and discuss deep $BR$ photometry of three star clusters in the SMC:
Kron~11, Kron~63 and NGC~121. We have derived radial density, surface brightness
profiles and LFs for these clusters. NGC~121 is a well known compact cluster, but
Kron~11 and Kron~63 are notably less prominent. We confirm that Kron~63 (the less
conspicuous of these latter two) is a real object, which resembles more a Galactic
than a Globular cluster. These basic conclusions are confirmed by the $\gamma$ value
of their profile fit, from a shallow exponent of 1.3 for Kron~63 giving it a looser
appearance in the sky, to the sharper NGC~121 cluster with a $\gamma$ value of 3.17.

From the analysis of deep LFs and CMDs we provide for the first time estimates of the
main parameters of Kron~11 and Kron~63. While Kron~11 is an intermediate age cluster,
Kron~63 is significatively younger. Interestingly, both clusters look coeval to the
surrounding stellar field.

We have confirmed the metallicity and age values previously estimated for NGC~121;
and found a clear mass segregation effect, a process that is expected to happen in a
cluster as old as it is.

\section*{Acknowledgments}
GB acknowledges support from the Chilean {\sl Centro de Astrof\'\i sica} FONDAP No.
15010003 and from CONICET (PIP~5970). EC and RAM acknowledge support by the Fondo
Nacional de Investigaci\'on Cient\'ifica y Tecnol\'ogica (proyecto No. 1050718,
Fondecyt) and by the Chilean {\sl Centro de Astrof\'\i sica} FONDAP No. 15010003.
CG and NN acknowledge support by the Instituto de Astrofísica de Canarias (P3-94)
and by the Ministry of Education and Research of the Kingdom of Spain (AYA2004-06343).

\end{document}